\documentclass[useAMS,usenatbib,letters]{mnras}

\usepackage{amsmath,amssymb}
\usepackage{bm}
\usepackage{mathtools}
\usepackage{graphicx}
\usepackage[usenames,dvipsnames]{color}
\usepackage{varioref}
\usepackage{float}
\usepackage[normalem]{ulem}
\newcommand{\stkout}[1]{\ifmmode\text{\sout{\ensuremath{#1}}}\else\sout{#1}\fi} 
\usepackage{rotating}
  \newcommand*{\swap}[2]{#2#1}
  \newcommand{\mkG}{\,\mu{\rm G}}
  \newcommand{\cmcube}{\,{\rm cm^{-3}}}
  \newcommand{\erg}{\,{\rm erg}}
  \newcommand{\g}{\,{\rm g}}
  \newcommand{\kms}{\,{\rm km\,s^{-1}}}
  \newcommand{\K}{\,{\rm K}}
  \newcommand{\kpc}{\,{\rm kpc}}
  \newcommand{\pc}{\,{\rm pc}}
  \newcommand{\yr}{\,{\rm yr}}
  \newcommand{\mnpr}{$\mathcal{P}_B(s)$}
   \newcommand{\rnpr}{$\mathcal{P}_b(s)$}
  \newcommand{\vpr}{$\mathcal{P}_{V}(s)$}

  \newcommand{\vect}[1]{{{\mbox{\boldmath $#1$}}}}
  \newcommand{\df}[1]{{\mathrm d#1}}
  \newcommand{\dfn}[2]{{\mathrm d^{#1}#2}}
  \newcommand{\PI}{{Paper\,I}} 				
  \newcommand{\PII}{{Paper\,II}} 				
  \newcommand{\rmp}{{Rev.\ Mod.\ Phys.}}
  \newcommand{\MFD}{{_\mathrm{MFD}}} 
    \newcommand{\FD}{{_\mathrm{FD}}} 

\title[Magnetic fields in the multi-phase ISM]{The distribution of mean and fluctuating magnetic fields in the multi-phase ISM}
\author[C.~C.~Evirgen, F.~A.~Gent, A.~Shukurov, A.~Fletcher and P.~Bushby]{
C.~C.~Evirgen$^{1}$\thanks{
E-mails:\,\,\,c.c.evirgen@newcastle.ac.uk;\,\,\,frederick.gent@aalto.fi; 
anvar.shukurov@newcastle.ac.uk;\,\,\,andrew.fletcher@newcastle.ac.uk;
paul.bushby@newcastle.ac.uk}, 
F.~A.~Gent$^{2}$, A.~Shukurov$^{1}$, A.~Fletcher$^{1}$ and P.~Bushby$^{1}$.\\
$^{1}$School of Mathematics and Statistics, Newcastle University, Newcastle 
upon Tyne, UK, NE1 7RU\\
$^{2}$ReSoLVE Centre of Excellence, Department of Computer Science, Aalto University, PO Box 15400, FI-00076 Aalto, Finland
}
\newcommand{\bvec}[1]{\boldsymbol{#1}}
\newcommand{\avg}[1]{\left<\bvec{#1}\right>_{l}}
\begin{document}
\date{Accepted -. Received -; in original form -}

\pagerange{\pageref{firstpage}--\pageref{lastpage}} \pubyear{2015}

\maketitle

\label{firstpage}

\begin{abstract}
We explore the effects of the multi-phase structure of the interstellar
medium (ISM) on galactic magnetic fields. 
Basing our analysis on compressible magnetohydrodynamic (MHD)
simulations of supernova-driven turbulence in the ISM, we investigate the 
properties of both the mean and fluctuating components of the 
magnetic field. 
We find that the mean magnetic field preferentially resides in the warm
phase and is generally absent from the hot phase.
The fluctuating magnetic field does not show such pronounced sensitivity to the 
multi-phase structure. 
\end{abstract}

\section{Introduction}

  The interstellar medium (ISM) has a complex, multi-phase structure. 
  However, very little is known 
  about the influence that this structure has upon galactic magnetic fields.  
  This is partly due to limitations in the observational techniques, but it should also be
  emphasised that galactic dynamo theory has been developed without any explicit reference to
  the multi-phase structure of the ISM \citep{BBMSS96,S07}.
  Further theoretical progress is needed to 
  aid the interpretation of observations. 
  
  Two types of dynamo operate in a typical spiral galaxy.  
  The mean-field (large-scale) dynamo produces a magnetic field that is ordered on a
  scale larger than the turbulent scale, $l_0\simeq50$--$100\pc$.
  This process relies on the differential rotation of
  galactic gaseous discs as well as helical turbulence in the ISM.
  The e-folding time of the large-scale magnetic field, $T\MFD$, is comparable to the turbulent
  magnetic diffusion time across the ionised gas layer, which is of the order of  
  $2.5\times10^8\yr$ near the Sun. 
  The other key dynamo mechanism is the fluctuation (small-scale) dynamo, in which
  local turbulent motions (which may, or may not, be helical) produce a disordered magnetic field
  that is structured on the scale of the flow  \citep[e.g.][]{ZRS90,BS05}.  
  The time scale at which these small-scale magnetic fields are amplified is of the
  order of the eddy turnover time of the turbulent flow,
  $T\FD\simeq l_0/v_0\ (\simeq10^7\yr$ in the warm phase near the Sun, assuming
  that the scale and speed of interstellar turbulence are $l_0=100\pc$ and 
  $v_0=10\kms$, respectively).
  Both types of dynamo mechanism amplify magnetic fields up to a strength of the order of a few microgauss,
  which corresponds to energy equipartition with the turbulence, $B_0\simeq(4\pi\rho v_0^2)^{1/2}$,  { where
  $\rho$ is the gas density}.
 
  Since the fractional volume occupied by the cold and molecular gas in the ISM is 
  negligible, it is likely that only the warm and hot phases affect 
  significantly dynamo action at the galactic scale.  { Therefore, here we
  focus on magnetic fields in the warm and hot diffuse gas phases.}
  The spatial scale of the mean magnetic field, of the order of $1\kpc$ or more, is
  comparable to or exceeds the typical size of the hot regions in the ISM.
  Furthermore it is replenished by the dynamo at a time scale longer than the residence 
  time of a parcel of hot gas within the gas layer, 
  $h/V_z\simeq5\times10^6\yr$, where $h\simeq500\pc$ is the scale height of the
  warm, partially ionised gas layer and $V_z\simeq100\kms$ is the vertical
  speed of the hot gas at the base of a galactic fountain or wind.
  Therefore, it seems plausible that the large-scale magnetic field should be
  mainly produced in the warm interstellar gas that remains in an average
  hydrostatic equilibrium within a relatively thin layer \citep{S07}.
  It is also important to note that, given the large volume fraction occupied by
  the warm phase, it is likely to form, on average, a simply connected 
  (percolating) volume in which the mean field can reside.
  On the other hand, the time scale of the mean-field dynamo is so much longer
  than the residence time of the hot gas in the warm layer that the dynamo
  might be controlled by ISM parameters averaged over time scales
  comparable to $T\MFD$; then the mean magnetic field would permeate both the
  warm and hot phases.
   { Thus, order of magnitude estimates alone do not provide us with sufficient information to 
  determine which phase of the ISM maintains the large-scale magnetic field.}

  The time scale of the fluctuation dynamo $T\FD$ also exceeds the residence
  time of the hot gas in the warm layer, but not by a wide margin.
  It is therefore plausible that the fluctuation dynamo is able to amplify the
  random magnetic field in the hot gas to the level of equipartition with the local
  turbulence only at a certain height above the galactic midplane, while the
  magnetic field strength in the hot gas near the midplane is significantly 
  below equipartition  { as it is produced from the field of the warm phase via expansion.}

  The structure of this Letter is as follows. In Section~\ref{sec:sims} we briefly 
  describe the numerical simulations which are the source of our data. The method 
  we use to define the magnetic field lines of the mean and fluctuating magnetic 
  field components is covered in Section~\ref{sec:methods}. In Section~\ref{sec:fld_lns}
  we investigate how the mean and fluctuating magnetic fields are connected to the 
  different phases of the ISM. The main conclusions are summarised in 
  Section~\ref{sec:conclusions}.
\section{Simulations of the multi-phase ISM}
\label{sec:sims}

  It is now possible to carry out magnetohydrodynamic (MHD) simulations of the ISM, including  
   { most of} the relevant physical processes \citep[e.g.][]{Korpi99a,Korpi99,AB05a,MacLow05,Gressel08,Piontek09,HJMBM12,
         BGE15,HSKHL15}. Our results  { use} the simulations of 
         supernova-driven turbulence in the multiphase ISM
  of \citet{GSFSM13} and \citet{GSSFM13}, subsequently referred to
  as \PI\, and \PII, respectively.
  The crucial point about these simulations is that the magnetic 
  field has not been imposed, but evolves dynamically under realistic physical
  conditions {, including the dynamo action} \citep[see also][]{Gressel08,BGE15}.
  The numerical model  { solves} the non-ideal MHD equations  {\citep[described in detail in][Section 3]{Gent12}}, 
  in a local box of $1\times1\kpc^2$ horizontally
  and $-1<z<1\kpc$ vertically
   { in size}, with the galactic midplane 
  at $z=0$.
  Gravity due to stellar mass and the dark halo follows \citet{Kuijken89}. 
  All models are subject to radiative cooling \citep{Sarazin87,Wolfire95},
  photoelectric heating \citep{Wolfire95} and other transport processes, which
  are necessary to support the multiphase structure. Local estimates for the differential rotation, supernova rate and 
  distribution, and column density are used \citep[see][]{F01}. 
  A  { nanogauss} seed magnetic field is amplified by dynamo action until it 
  saturates with a typical magnetic field strength of a few  { microgauss}.

  We follow \citet{Gent12} in defining the three phases of the ISM in 
  terms of  { specific entropy $s$,  
  expressed as
  \begin{equation}
    \label{eq:ent}
    s = c_V\left[\ln(T/T_0)-(\gamma-1)\ln(\rho/\rho_0)\right],
  \end{equation}
  where $\rho$ (base unit, $\rho_0=1\g\cmcube$) and $T$ (base unit, $T_0=1\K$)
  denote density and temperature, respectively, $c_V$ is the specific heat
  capacity at constant volume, and the adiabatic index is $\gamma=5/3$.
  Using Eq.\,\eqref{eq:ent} and in units of $10^8$ erg g$^{-1}$ K$^{-1}$, 
  the cold phase is defined as $s<4.4$, 
  the warm phase as $4.4<s<23.2$, 
  and the hot phase as $s>23.2$.
  The phases of the ISM can also be defined according to temperature and density.}
  The phase definitions are listed in Table\,\ref{tab:mphase} together with the typical
  temperature and density within these entropy  { ranges}.
\begin{table}
  \caption{ 
   { Parameters} of the ISM phases:
   { specific entropy $s$ [$10^8$ erg g$^{-1}$ K$^{-1}$], defined in Eq.\,\eqref{eq:ent}, 
  temperature $T$ [K] and density $\rho$ [g cm$^{-3}$].
  The phases are in pressure equilibrium, 
  with total pressure log-normally distributed about
  $10^{-12.5}$~dyn cm$^{-2}$ \citep[][Figure\,5.11d]{Gent12}.}}
  \label{tab:mphase}
  \begin{tabular}{lccc}
  \hline  { ISM phase} & Cold & Warm & Hot \\
    \hline
       $s$ &    $s<4.4$      & $4.4     < s  <23.2$        & $s>23.2$       \\
       $T$ &    $T<500$      & $500     < T  <5\cdot10^5$ & $T>5\cdot10^5$\\
    $\rho$ & $\rho>10^{-24}$ & $10^{-26}<\rho<10^{-24}$    & $\rho<10^{-26}$\\
    \hline   
  \end{tabular}
  
\end{table}

  We consider volume and time averages of physical variables from 23 snapshots from a nonlinear 
  MHD model that has twice the galactic rotation rate of the solar neighbourhood. 
  Integrating MHD models to attain dynamo saturation is computationally expensive 
  (even the most efficient dynamo from \PII\ took over 1\,Gyr to reach saturation). 
  Our choice of rotation rate is a pragmatic one, designed to optimise the efficiency of 
  the dynamo. We will consider models with lower rotation rates in future work. 
  To illustrate the difference that a magnetic field makes to the phase-structure of the 
  ISM we also consider snapshots taken from the kinematic phase of the dynamo, during 
  which the field is too weak to influence its surroundings.
\section{The mean and fluctuating magnetic fields}\label{sec:methods}
  The decomposition of the magnetic field into mean and fluctuating (random)
  parts follows the method described in \PII.
  Volume averaging with a Gaussian kernel $G_{l}(\bvec{x}-\bvec{x}')$
   { of a scale $l$} is used to  { split} the magnetic field $\bvec{B}$ 
  into mean, $\bvec{B}_{l}$, and random, $\bvec{b}_{l}$,  { parts}:
  \begin{equation}
    \bvec{B} = \bvec{B}_{l}+\bvec{b}_{l},\qquad \bvec{B}_{l}
             = \avg{B},
    \label{eq:decomp} 
  \end{equation}
  where  { angular brackets denote an average} calculated as
   \begin{align}
    \label{eq:gauss}
    \avg{B}(\bvec{x})&=
         \int_{V}\bvec{B}(\bvec{x}')G_{l}(\bvec{x}-\bvec{x}')\dfn{3}{\bvec{x}'},\\
    G_{l}(\bvec{x})&=\left(2\pi l^2\right)^{-3/2}
                     \exp\left[-\bvec{x}^2/(2l^2)\right],\nonumber
  \end{align} 
  where $l\approx50$\,pc is half the integral scale of the turbulent motions
   { in the numerical model (see \PII\ for further details)}.
  Preliminary analysis does not show significant sensitivity of the mean or
  random field to variations in $l$ within the range $30<l<100$\,pc.
  
  Given a magnetic field, $\bvec{B}(\bvec{x})$, in Cartesian coordinates, its
  integral (field) lines are described by
  \begin{equation}
    \dfrac{\df{x}}{B_{x}}=\dfrac{\df{y}}{B_{y}}
                     =\dfrac{\df{z}}{B_{z}}
                     =\dfrac{\df{r}}{|\bvec{B}|},
    \label{f_lines_eqn}
  \end{equation}
  where $\df{r}$ is  { the line element measured along the line}. 
  We obtain the integral lines for both the mean and fluctuating magnetic fields by integrating these equations,
  using a  { fourth-}order Runge-Kutta scheme, applying linear interpolation 
   { between the} grid points.
  
  Our aim is to determine  { whether} the mean and fluctuating magnetic field {s  
  are predominantly located} in specific phases of the ISM. However, it is 
  not straightforward to find a robust quantitative measure for this. 
   { The spatial distribution} of magnetic energy density is biased towards the cold, dense gas that
  occupies a negligible fraction of the volume. We suggest a different approach, based on a comparison 
  of  { the statistical properties of specific entropy} along field lines with  { those} in the entire volume.
  If a magnetic field does not prefer to reside in any  {particular} phase, the probability density function (PDF) of
   { specific} entropy sampled along the field lines should be the same as the volume PDF.
  Conversely, if a magnetic field is sensitive to the multi-phase structure,
  the difference between the  { field-line and volume PDFs of specific entropy} will highlight the
  entropy interval(s), and thus the phase(s) where differences arise.
   
\section{Magnetic fields in the multi-phase structure}\label{sec:fld_lns}

\begin{figure}
    \centering 
    {\includegraphics[width=1.02\linewidth]{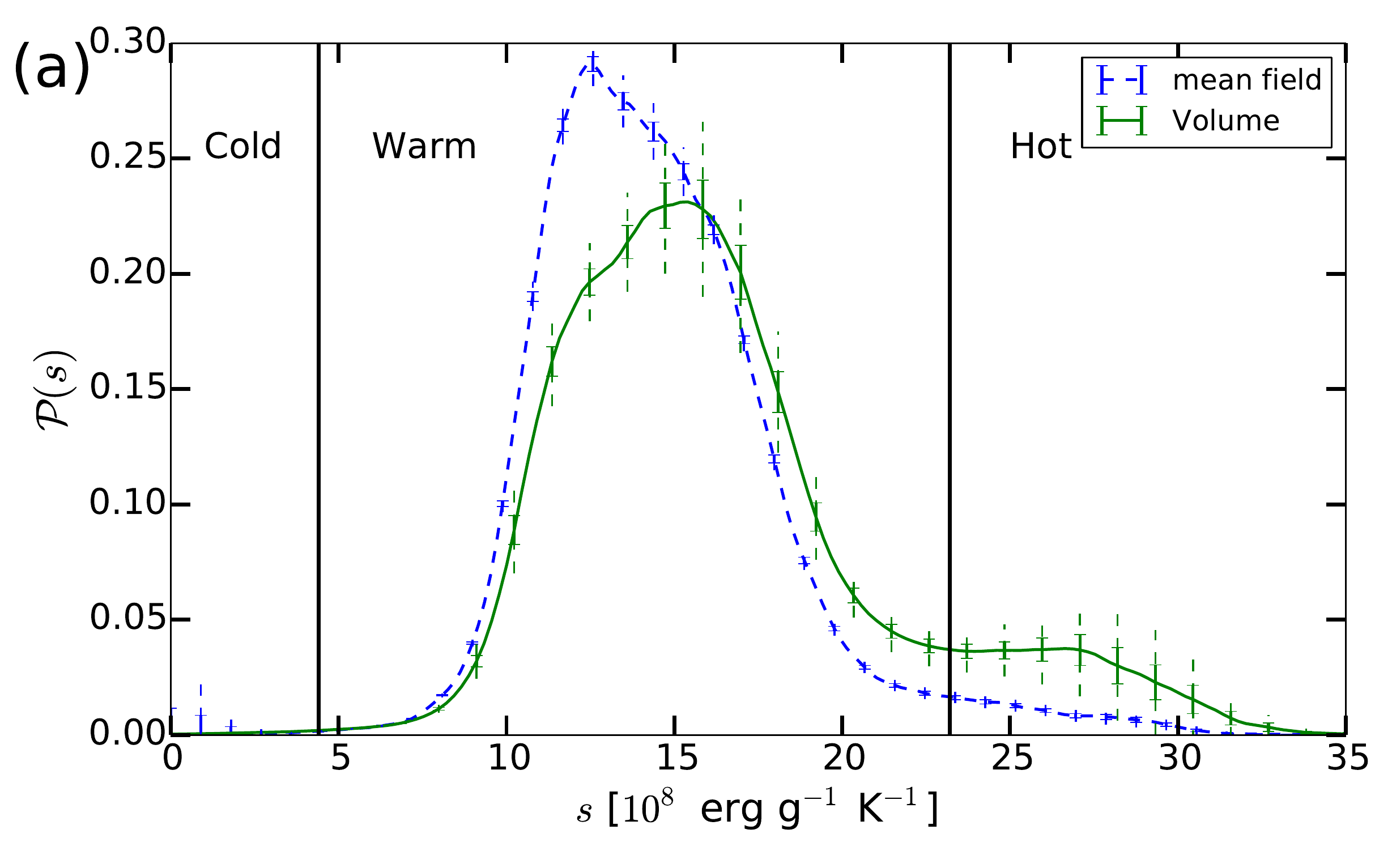}}
    {\includegraphics[width=1.02\linewidth]{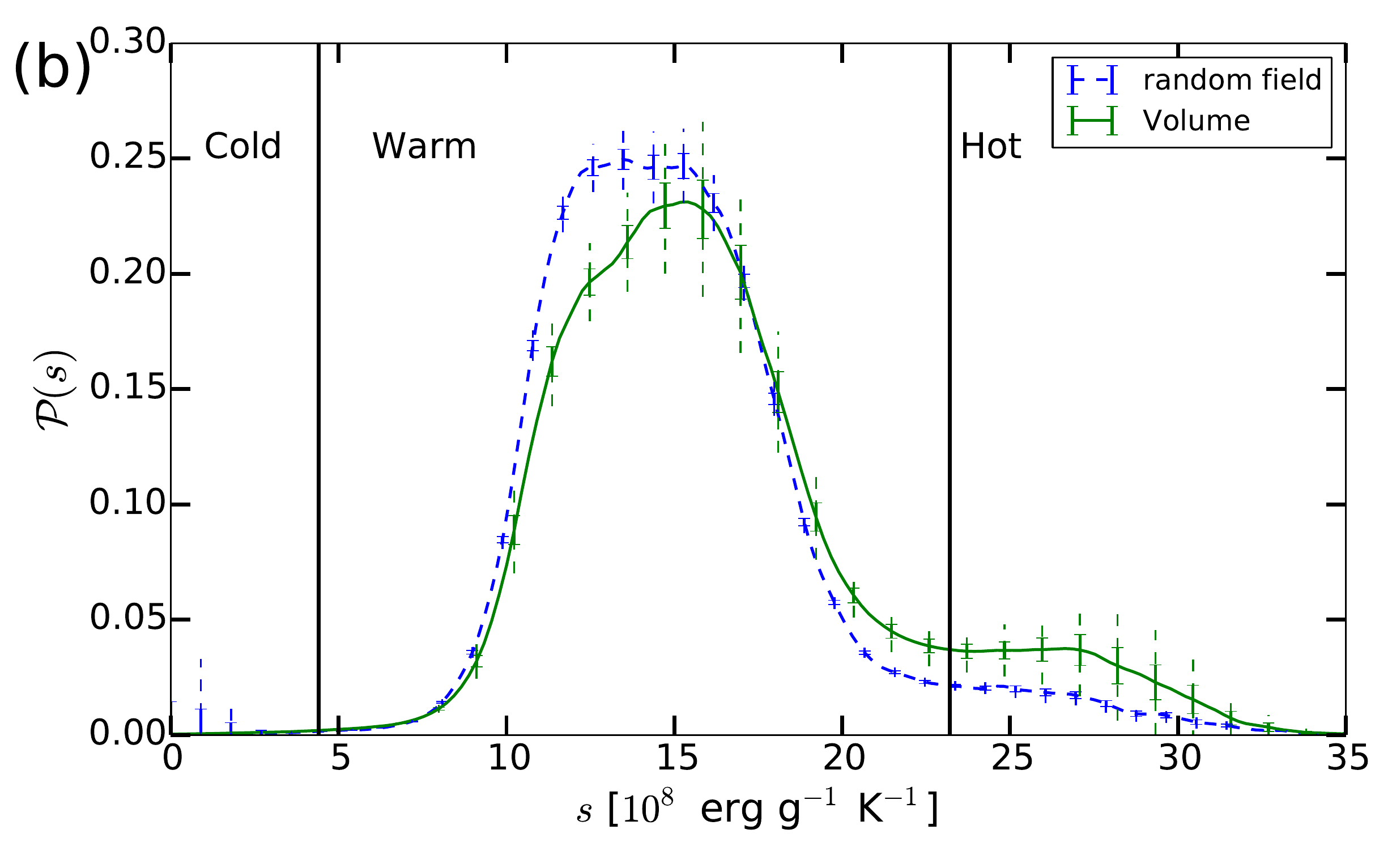}}
    \caption{
       { The probability density functions} (PDFs) of  { specific} entropy  { in the whole computational volume (solid) and 
      sampled} along the integral lines   {(dashed)} of the
      {\textbf{(a)}}~mean 
      and {\textbf{(b)}}~random magnetic fields. 
      Vertical lines show the boundaries between the cold, warm and hot  { ISM} phases.
      \label{fig:ent_pdf_mn}
    }
  \end{figure}

  Figure\,\ref{fig:ent_pdf_mn}(a) compares \vpr, the volume-sampled  { specific} entropy PDF, with \mnpr, the  { corresponding} PDF 
  sampled along the mean magnetic field lines. These plots indicate that the mean magnetic field tends to favour the low entropy zone of the warm phase; the peak of \mnpr\ is located at $s=12$  {(the specific entropy is expressed here, and elsewhere in the text, in units of $10^8\erg\g^{-1}\K^{-1}$)}, whereas the corresponding peak in \vpr\ is located at $s=15$. For $18\lesssim s<23$, \mnpr\ is systematically lower than \vpr. Furthermore, for entropy values in the range $s>23$, \mnpr\ is significantly lower than \vpr, which suggests that the mean field  { avoids} the hot gas. Figure~\ref{fig:ent_pdf_mn}(b) shows a comparison between \vpr\ and \rnpr, which is the PDF  { of specific entropy along} the fluctuating (random) magnetic field. The differences between these curves are less dramatic than those shown in 
  Figure\,\ref{fig:ent_pdf_mn}(a). Whilst \rnpr\ systematically has a higher probability density than \vpr\ for $s<15$, the
  difference is clearly smaller than for \mnpr. 
  The random field component is suppressed to some extent in the hot phase, but this is
  less pronounced than it is for the mean magnetic field.

  \begin{figure*}
    \centering
    {\includegraphics[width=0.30\linewidth,
                   trim=0.5cm 0.0cm 0.5cm 0.0cm, 
                   clip=true, 
    ]{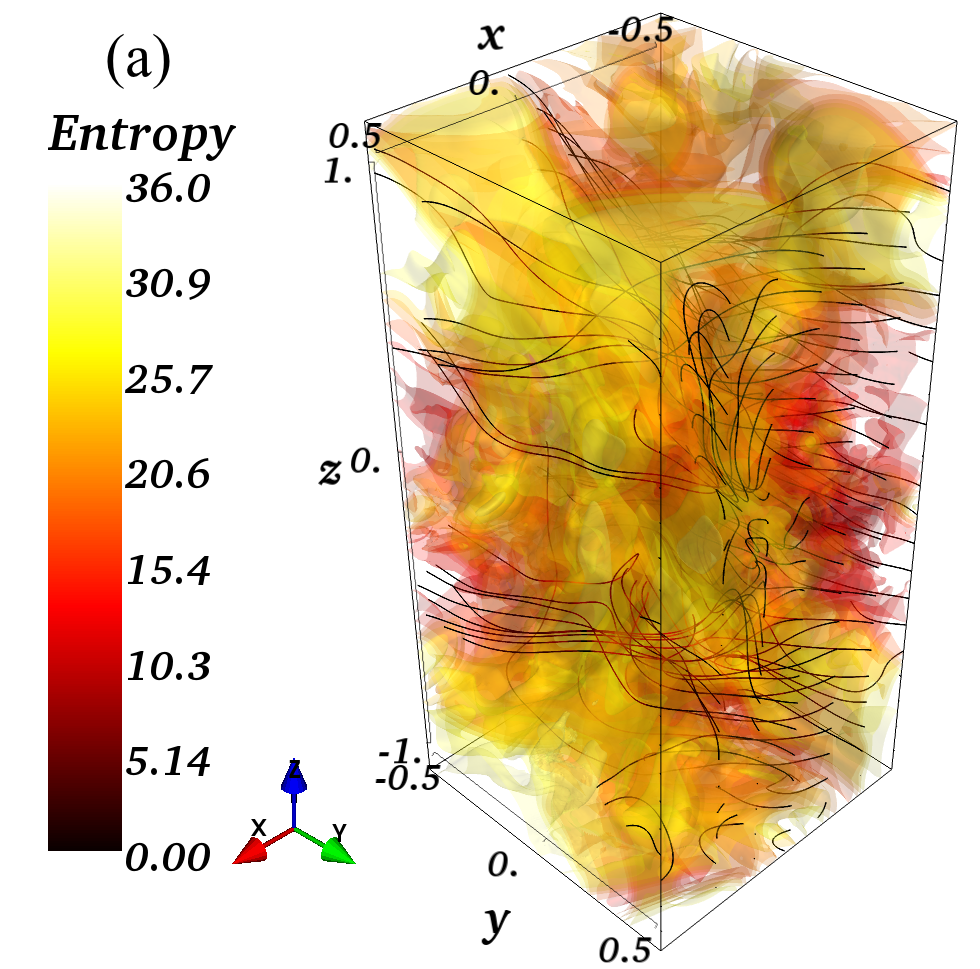}}
      \hfill    
    {\includegraphics[width=0.23\linewidth,
                   trim=1.7cm 0.0cm 1.5cm 0.0cm, 
                   clip=true, 
    ]{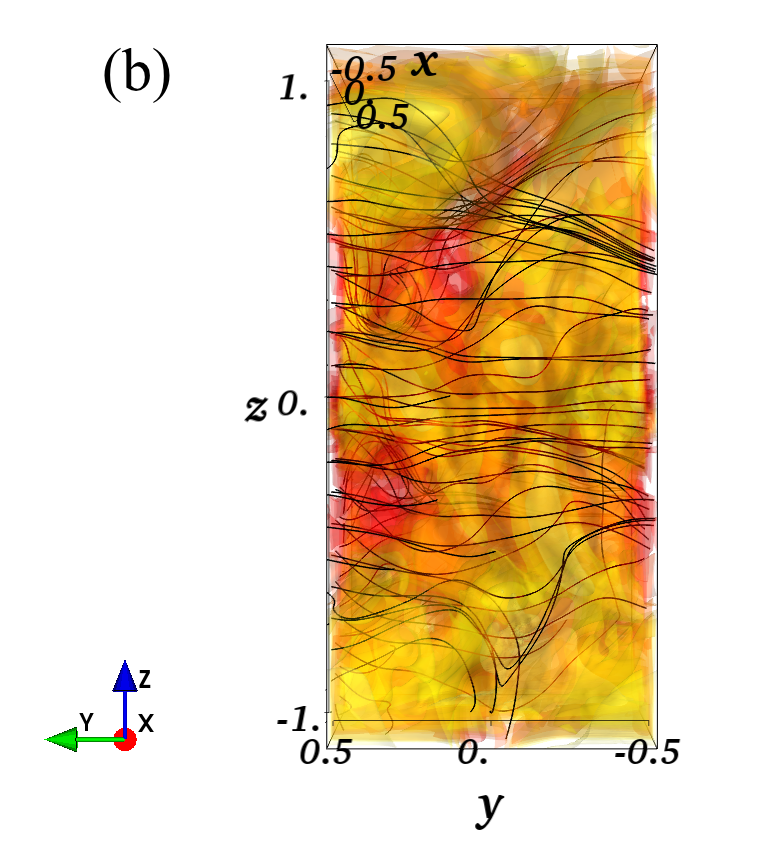}}
      \hfill  
    {\includegraphics[width=0.24\linewidth,
                   trim=1.0cm 0.0cm 0.5cm 0.0cm, 
                   clip=true, 
    ]{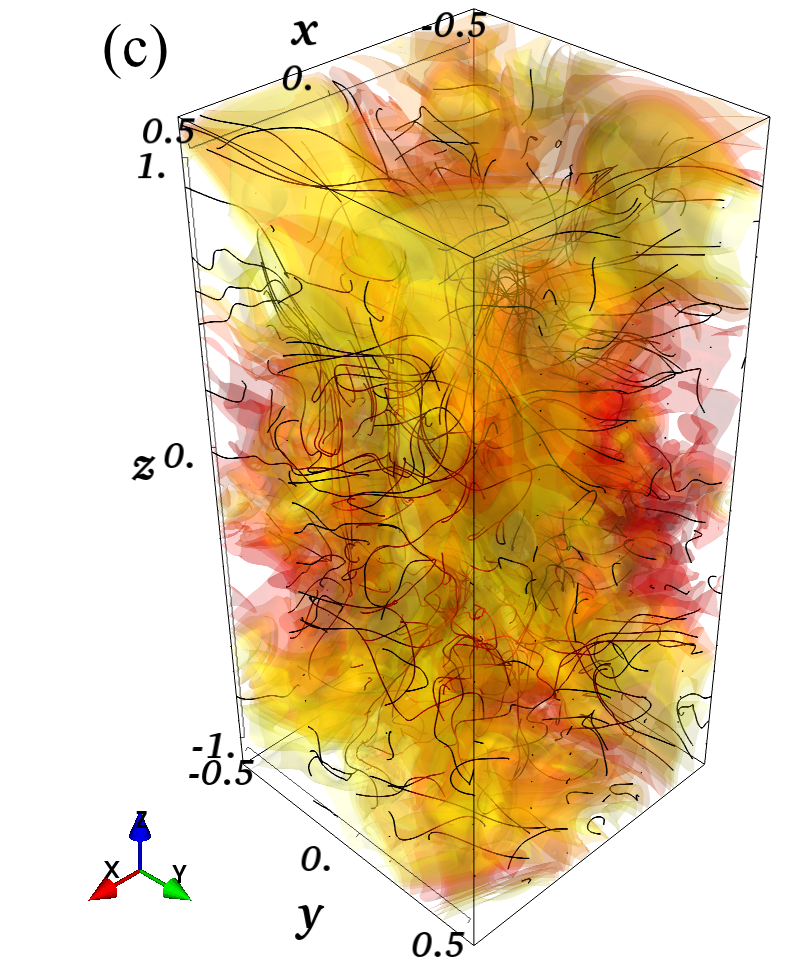}}
      \hfill      
    {\includegraphics[width=0.21\linewidth,
                   trim=1.7cm 0.0cm 1.5cm 0.0cm, 
                   clip=true, 
    ]{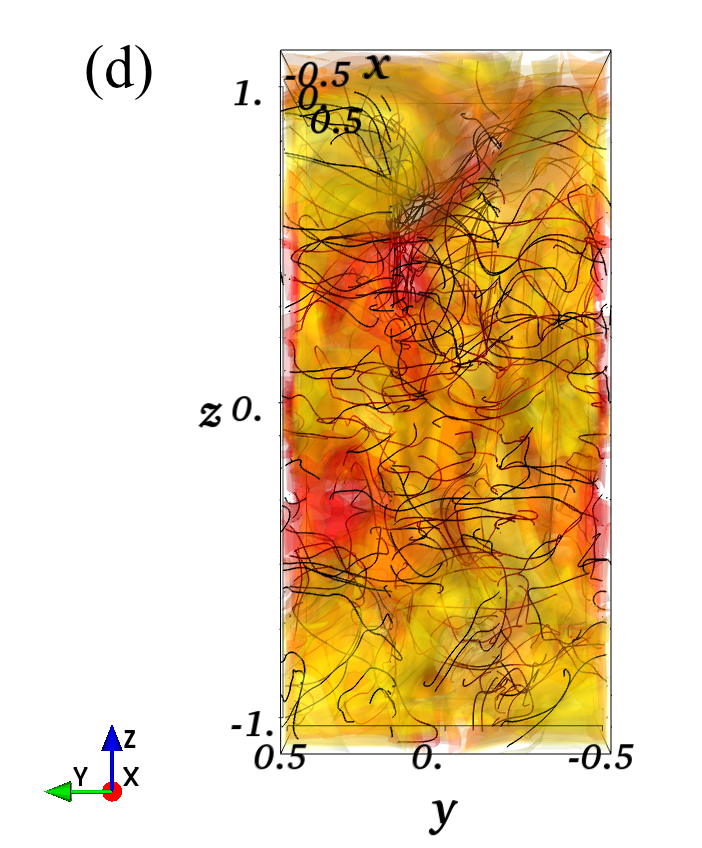}}
    \caption{
      3D rendering of magnetic field lines (black) in the 
      simulated ISM, with  { the specific} entropy  { of the gas} in the background  { (colour, in the units of
      $10^8\erg\g^{-1}\K^{-1}$)}. Panels \textbf{(a)} and \textbf{(b)} show the 
      mean-field lines and panels \textbf{(c)} and \textbf{(d)} the random field lines.
      Panels \textbf{(a)} and \textbf{(c)} give an isometric view, and panels \textbf{(b)}
      and \textbf{(d)} show a view through the $(y,z)$ plane.
       { Cartesian coordinates $(x,y,z)$ locally correspond to the cylindrical polar coordinates $(r,\phi,z)$
      with the $z$-axis aligned with the angular velocity of galactic rotation}.
      \label{fig:vis_td}
    }
  \end{figure*}

Figure~\ref{fig:vis_td} uses a single snapshot in the nonlinear regime to give an alternative view of these results. In panel~(a), there is a 
large column (chimney) of hot, high-entropy gas spanning the domain horizontally and vertically, from which mean magnetic field lines appear to be absent. This is consistent with the PDFs shown in Figure~\ref{fig:ent_pdf_mn}, further reinforcing the idea that the mean magnetic field is sensitive to the multi-phase structure. Panel~(b) shows that the mean magnetic field, where it is found, tends to be  { approximately} aligned with the  { azimuthal ($y$) direction (as it is affected by the velocity shear)}.  Panels (c) and (d) show the random (fluctuating) magnetic field in the same snapshot. As expected, the field lines do not appear to have a preferred direction. In addition, the random magnetic field lines do not appear to avoid the column of hot gas in the same way as the mean field. Thus, the random magnetic field appears to be less sensitive to the multi-phase structure.  

  \begin{figure}
    \centering 
    \includegraphics[width=1.02\linewidth,trim=1.5cm 0.0cm 0.5cm 1.5cm,clip=true]{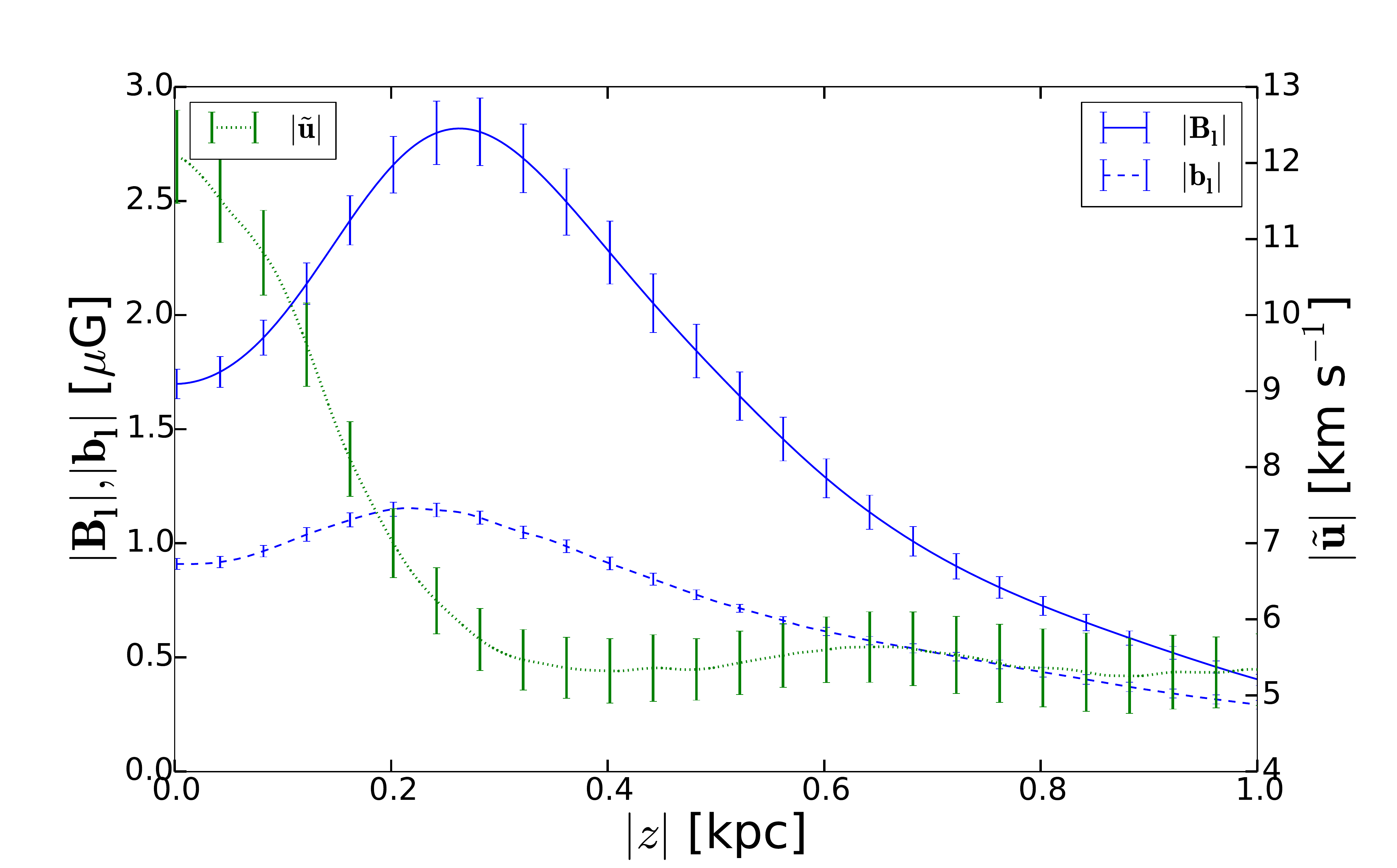}
    \caption{
      Horizontal averages of mean magnetic field strength $|\bvec{B}_l|$  {(solid lines)}, random magnetic field strength $|\bvec{b}_l|$  {(dashed lines)} and random velocity  
      { $|\tilde{\bvec{u}}|$ (dotted lines)}, shown as functions of distance from the midplane.
      \label{fig:u2bz}
    }
  \end{figure} 

 { It is plausible that the relative reduction in the strength of the mean magnetic field in the hot gas is explained by the rapid expansion of hot gas bubbles. Furthermore, the hot gas is removed from the galactic disc over a time scale significantly shorter than the mean-field dynamo time scale.} Figure~\ref{fig:u2bz} shows that the mean and random
  magnetic fields reach their maximum amplitude away from the midplane,
   { just outside the layer $|z|\lesssim0.2\kpc$ where most of the supernovae are located}.
  We note that the random magnetic field strength remains
  of order $1\mkG$ for $|z|<0.2\kpc$, whilst the mean magnetic field increases
  from  {$1.7\mkG$} at the midplane to  {$2.8\mkG$} at 
   {$|z|=0.3\kpc$}. The mean field strength at the midplane is remarkably consistent with the observed estimate of 
  of \citet{RK89}.  { However, we note that the strength of the random magnetic field in our simulations
  is significantly lower than the $5\mkG$ observed in the solar vicinity of the Milky Way \citep{BBMSS96,Haverkorn15}.}
   { Whilst we do not believe that this discrepancy affects our main conclusions regarding the distribution of the field across the ISM phases, t}he reason for this difference is not obvious. It may indicate that the fluctuation dynamo (which directly generates small scale field) is less efficient than it should be, so that the  { simulated} random field is due primarily to the tangling of mean  { magnetic} field lines by the turbulent velocity field. Another possibility is the (implicit) use of longer averaging scales in the interpretation of the observations. Our domain side of $1\kpc$ limits the smoothing scale that we can apply. However, these possibilities are speculative and more work is required to properly understand the relatively weak random field in the simulations.

Figure\,\ref{fig:u2bz}  { also shows} the random velocity  {$|\tilde{\bvec{u}}|$
 (defined in the rotating frame with the mean \emph{vertical} flow deducted) for which} there is a
 local maximum at the midplane, where the supernovae dominate the dynamics.
  Away from the midplane, the random velocity decreases rapidly reaching a minimum value at approximately $|z|\sim0.4$~kpc, 
  where the mean magnetic field is strong. At larger values of $|z|$, the amplitudes of the mean and fluctuating components of 
  the magnetic field both decrease with increasing distance away from the midplane. In this region, the mean magnetic field 
  strength decreases from its maximum, $2.8\,\mu$G, to $0.4\,\mu$G.
  The decrease in random magnetic field strength is more modest
  (1 to $0.3\,\mu$G). The variation of the magnetic field with $|z|$ suggests that the most efficient dynamo action is confined primarily to regions within a few hundred parsecs of the midplane.
  
\begin{figure}
    \centering 
    \includegraphics[width=1.02\linewidth]{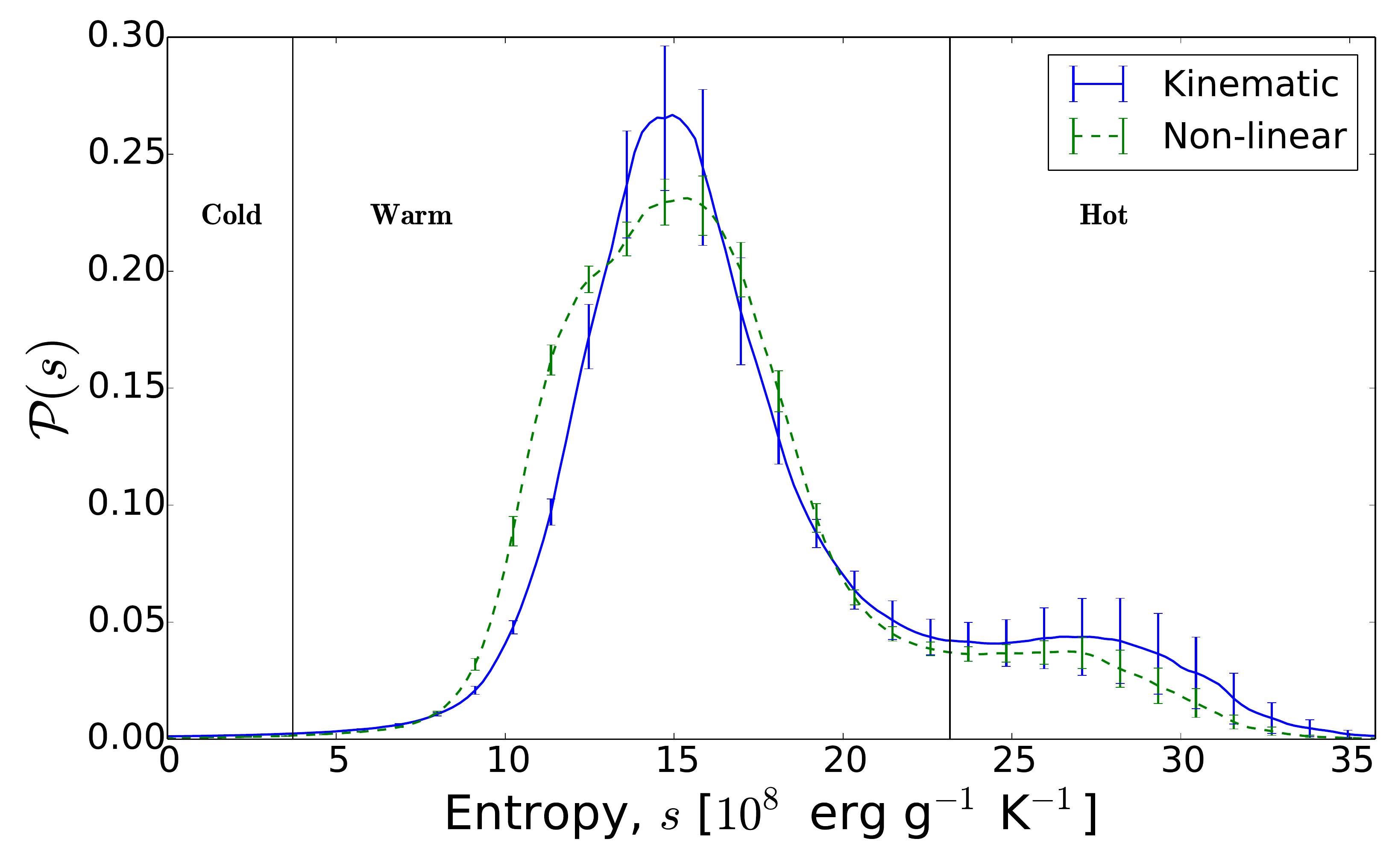}
    \caption{
       { Probability density of specific entropy in the computational domain for} the kinematic ( { solid}) and nonlinear ( { dashed}) state of the dynamo. Vertical lines show the boundaries between the cold and warm ISM phases and the warm and hot phases. 
      \label{fig:model_comp}
    }
  \end{figure}

  Figure\,\ref{fig:model_comp} displays the PDFs of  { specific} entropy  { in the whole computational domain} during both the early kinematic and nonlinear (saturated) 
  dynamo stages.
  There is a difference in the distribution of entropy between the ISM with a dynamically-insignificant (i.e. kinematic) magnetic field and the ISM with a dynamo-generated magnetic field that has saturated. 
  Whilst the modal probabilities of the warm and hot phase are
  similar, the shape of the distribution is
  different. In the case of a saturated dynamo,
  the PDF is wider in the warm phase, and has a region of higher probability
  density in $10<s<12$.
  In addition, saturation of the dynamo leads to a consistent reduction of
  probability density for the higher entropy gas with $s>20$.
  Even though there is clear evidence for the existence of a warm and hot phase
  the entropy distributions within the phases change as the magnetic
  field grows. 

   { Further insight into the mean-field dynamo mechanism can be gained by examining the electromotive force (EMF). Denoted by $\vect{\cal E}$, the EMF can be expressed as a sum of its mean, $\vect{\cal E}_l$, and 
   fluctuating, $\vect{\cal E}^\prime$, parts. These are calculated as follows:
  \begin{equation*}
    \vect{\cal{E}} = \vect{u}\times\vect{B}, \qquad \vect{\cal{E}}_{l}
             = \left< \vect{u}\times\vect{B}\right>_l,\qquad \vect{\cal{E}'} = \vect{\cal{E}}-\vect{\cal{E}}_l,
    \label{eq:decomp} 
  \end{equation*}
  where $\vect{u}$ denotes the total velocity field in the rotating frame.
   Summary statistics for the mean and fluctuating EMF are given in Table\,\ref{tab:emf}. These values indicate that the 
   mean EMF in the warm phase is approximately twice as strong as it is in the hot phase, which supports the idea that dynamo action in the mean field is strongest in the warm phase. The warm and hot phases have similar values for the fluctuating part of the EMF. }

\begin{table}
  \centering
 {
  \caption{
     Averages of the mean and fluctuating EMF strength over the volumes $\cal V$ occupied 
     by the warm or hot phases.   
    ${\cal E_\ell = \left<|\vect{\cal E}_\ell|\right>_{\cal V}}$ and 
    ${\cal E^\prime = \left<|\vect{\cal E^\prime}|\right>_{\cal V}}$, with standard 
      deviation denoted by $\sigma_\ell$ and $\sigma^\prime$\!, 
      respectively [G km s$^{-1}$]. 
  }
  \label{tab:emf}
  \begin{tabular}{lcccc}
  \hline
   & $\cal E_\ell$  & $\sigma_\ell$ & $\cal E^\prime$ & $\sigma^\prime$\\
  \hline
  Warm & 1.12 & 0.91 & 1.01 & 1.14\\
  Hot & 0.65 & 0.72 & 1.03 & 1.91 \\
  \hline
  \end{tabular}}
\end{table}

\section{Conclusions and discussion}\label{sec:conclusions} 
We have shown that the mean magnetic field is sensitive to the multiphase structure of the ISM. 
Our PDF analysis indicates that it resides preferentially in the lower entropy region of the warm phase, 
particularly in the layer $0.2<|z|<0.4$~kpc, avoiding regions of hotter gas. Given the presence of the velocity 
shear, it is unsurprising that this mean field tends to be aligned with the $y$-coordinate (i.e. the azimuthal direction) in our model. 
The random magnetic field appears to be less strongly influenced by the multiphase structure. As functions of distance from the midplane ($z=0$), the mean and random magnetic field strengths peak at $|z|=300$pc and $|z|=200$pc, respectively.   

 The marginal preference of the fluctuating field for low entropy regions 
  of the warm phase is likely due to generation of the random field by tangling of the mean field produced by the
  large-scale dynamo.
  Small-scale dynamo action may not be fully resolved with  { the grid
  resolution of $4$~pc in these simuations}, and so may be less efficient than it should be, but this interpretation is speculative. Separating the two different mechanisms,
  by which the random field can be produced is subtle and difficult; we shall return to this problem in subsequent work that examines how galactic dynamos saturate in the multi-phase ISM.
  
There is an increasing fractional volume of gas within the warm
phase, as the mean magnetic field grows and saturates.
Whilst it was expected that the magnetic field preferentially resides in
 the warm phase, this result suggests that dynamo action actively changes
 the volume entropy distribution, and thus the multi-phase structure of the ISM.
 This raises a significant question: does the magnetic field 
  preferentially reside in the warm phase, or does it adapt the multi-phase
  structure, in order to create a hospitable environment for dynamo action?
  In other words, how does the multi-phase structure change as the ISM becomes magnetised?
  We will discuss these questions, which can have important consequences for galactic evolution, in future work.

\section*{Acknowledgements}
  AS is grateful to Carl Heiles, Richard Crutcher and Thomas Troland for
  useful discussions of magnetic fields in the hot interstellar gas.
  FAG acknowledges financial support of the Grand Challenge project SNDYN,
  CSC-IT Center for Science Ltd. (Finland) and the Academy of Finland 
  Project 272157.
  AS, AF and PB were supported by the Leverhulme Trust 
  Grant RPG-2014-427 and STFC Grant ST/N000900/1 (Project 2).
  CCE was supported by the RAS and Nuffield Foundation with an RAS Undergraduate 
  Bursary, entitled ``Magnetic Fields and Turbulence in the Multi-phase 
  Interstellar Medium''.
   {  We would also like to thank the referee for useful comments and suggestions.}

  \bibliographystyle{mnras}     
  \bibliography{refs}

\label{lastpage}

\end{document}